\documentclass[11pt,a4paper,onecolumn]{article2}
\usepackage{graphicx}
\usepackage{subfigure}
\usepackage{caption2}
\usepackage{fancyhdr}
\begin{document}

\input epsf
\epsfverbosetrue

\setcounter{page}{1}
\pagenumbering{arabic}

\vspace{66pt}
\begin{center}
{\bf
ANALYSIS OF THE JINR p(660 MeV) + $^{129}$I, $^{237}$Np, AND $^{241}$Am
MEASUREMENTS WITH ELEVEN DIFFERENT MODELS
}\\
\vspace{33pt}
{\bf
S.~G.~Mashnik$^{1}$,
V.~S.~Pronskikh$^{2}$,
J.~Adam$^{2,3}$,
A.~Balabekyan$^{2,4}$,
V.~S.~Barashenkov$^{2}$,
V.~P.~Filinova$^{2}$,
A.~A.~Solnyshkin$^{2}$,
V.~M.~Tsoupko-Sitnikov$^{2}$,
R.~Brandt$^{5}$,
R.~Odoj$^{6}$,
A.~J. Sierk$^{1}$,
R.~E.~Prael$^{1}$,
K.~K.~Gudima$^{7}$,
M.~I.~Baznat$^{7}$
}

\vspace{-1mm}
$^1$Los Alamos National Laboratory, Los Alamos, New Mexico 87544, USA\\
$^{2}${\itshape Joint Institute for Nuclear Research, Dubna, 141980 Russia}\\
$^{3}${\itshape
 Institute for Nuclear Physics, Academy of Science of Czech Republic,
{\v R}e{\v z}, Czech Republic}\\
$^{4}${\itshape
 Yerevan State University, Republic of Armenia}\\
$^{5}${\itshape Institute of Nuclear Chemistry, Philipps University,
Marburg, Germany}\\
$^{6}${\itshape Forschungszentrum Julich, Germany}\\
$^{7}${\itshape Institute of Applied Physics, Academy of Science of
Moldova, Chisinau, Moldova}

\end{center}
\vspace{33pt}
\begin{center}
{\bf Abstract}
\end{center}
Isotopically enriched $^{129}$I  (85\% $^{129}$I and 15\%
$^{127}$I), $^{237}$Np, and $^{241}$Am targets were irradiated
with a beam of 660-MeV protons at the JINR DLNP Phasotron and
cross sections of formation of 207 residual products (74 from
$^{129}$I, 53 from $^{237}$Np, and 80 from $^{241}$Am)
 were determined using the $\gamma$-spectrometry method.
Here, we analyze all these data using eleven different models, realized in
eight codes: LAHET (Bertini, ISABEL,
INCL+ABLA, and INCL+RAL options),
CASCADE, CEM95, CEM2k, LAQGSM+GEM2, CEM2k+GEM2,
LAQGSM+GEMINI, and CEM2k+GEMINI,
in order to validate the tested models against the experimental data
and to understand better the mechanisms for production of residual nuclei.
The agreement of different models with the data varies quite a bit.
We find that most of the codes are fairly reliable in predicting
cross sections for nuclides not too far away in mass from the targets,
but differ greatly in the deep spallation, fission, and fragmentation
regions.
None of the codes tested here except GEMINI allow fission of nuclei
as light as iodine, therefore the best agreement with the $^{129}$I
data, especially in the A=40-90 region, is shown by the codes CEM2k and
LAQGSM when they are merged with GEMINI. At the same time, GEMINI is
not yet very reliable for an accurate description of actinides and
the $^{237}$Np and $^{241}$Am data are reproduced better by
LAHET (Bertini, ISABEL, or INCL+RAL/ABLA options),
and by CEM2k and LAQGSM merged with GEM2
and not so well when using GEMINI. We conclude that none of the
codes tested here are able to reproduce well all these data
and all of them need to be further improved; development of a better
universal evaporation/fission model should be of a high priority.

\newpage
\noindent{\bf Introduction}

\indent
Interest in  the physics of transmutation
(i.e., conversion into stable isotopes as a result of nuclear reactions)
of actinides and fission products  produced at nuclear power stations
has increased significantly during the last decade.
Estimations made by different groups \cite{1,2}
show that the radiation risk
of the spent nuclear fuel due to its possible leakage from deep underground
storage systems
after its transmutation is about the same as of
the uranium ore after 1000 years of storage, that is significantly
shorter than $5 \times 10^6$ years necessary to store the same spent fuel
without transmutation to decrease its risk to a similar level.

\indent
Analysis of the radiation hazard of the spent fuel showed that after
the extraction of
the uranium-plutonium group of elements and such
fission products like $^{90}$Sr, $^{137}$Cs, and $^{129}$I, the highest hazard
comes from $^{237}$Np and $^{241}$Am \cite{3}.
At that, $^{241}$Am $(T_{1/2} = 432.2$ years) contributes the most to the
radioactivity, while $^{237}$Np $(T_{1/2} = 2.144 \times 10^6$ years)
is dangerous because of its high concentration in the spent fuel
and its high migration
ability in the biosphere, that increases the probability
of its penetration into human
body through the food chain \cite{4}.

\indent
Investigation of $^{237}$Np and $^{241}$Am transmutation dynamics
in the flow of thermal neutrons of different densities shows that the
higher the density of neutrons, the smaller the number of different
actinides noticeably contributing to the radioactivity of wastes \cite{4}.
To solve the problem of transmutation, high-current proton accelerators
can be used to produce neutron fluxes of $\sim 10^{17}$ cm$^{-2}$c$^{-1}$
for transmutation purposes. In some recent publications, both transmutation of
actinides by thermal neutron irradiation and their spallation and fission
with the proton and ion beams are investigated \cite{5}.

\indent
Hadron-nucleus event generators are the basis for calculations of the
Accelerator Driven System (ADS) setups, their targets, and the blanket effect.
Such calculations are done using models of different accuracy. The best
test for different models and codes used in such applications is to
compare calculated and experimental yields of the residual nuclei from
reactions of interest. From experimental point of view, determination of the
independent cross-section for yields of short-lived nuclear products from
mono-isotope targets is the most important for such comparisons \cite{6}.
Experimental cross-sections for residual nuclei in radioactive $^{129}$I,
$^{237}$Np, and $^{241}$Am targets are undeniably important for the projects
of transmutation of nuclear wastes in a direct proton beam.
Measurements of the yield of residual nuclei from $^{237}$Np,  $^{241}$Am,
and $^{129}$I (85\% $^{129}$I and 15\% $^{127}$I) targets were recently
performed at the JINR Phasotron with proton beams of 660 MeV
\cite{7,8}.
In the present work, we analyze these measurements
with eleven models implemented in several event generators and transport
codes used in different nuclear applications, to
test these models against the experimental data and
with a hope to understand better mechanisms of nuclear reactions
and ways to improve the models and codes.

\noindent{\bf Results}

The  $^{237}$Np and  $^{241}$Am experimental
data are published in tabulated form in Ref.~\cite{7},
while the  $^{129}$I data are tabulated in \cite{8}. Details
on the measurements may be found in \cite{7,8} and we do not
discuss them here.

We analyzed all the measured data using eleven models, contained in
eight transport codes and  event generators.
Namely, we calculated the reactions with the LAHET3 version \cite{LAHET3}
of the transport code LAHET \cite{LAHET}
using the Bertini \cite{Bertini} and ISABEL \cite{ISABEL}
IntraNuclear-Cascade (INC) models  merged with the
Dresner evaporation model \cite{Dresner} and the Atchison
fission model (RAL) \cite{RAL},
and using the Liege INC code by Cugnon {\it et al.} INCL \cite{INCL}
merged in LAHET3 with the ABLA \cite{ABLA} and with
Dresner \cite{Dresner} (+ Atchison \cite{RAL}) evaporation
(+ fission) models,
with the Dubna transport code CASCADE \cite{CASCADE},
with versions of the Cascade-Exciton Model (CEM)
\cite{CEM} as realized in the codes CEM95 \cite{CEM95} and
CEM2k \cite{CEM2k}, with CEM2k merged \cite{Mashnik02a}-\cite{fitaf}
with the Generalized Evaporation/fission Model code GEM2 by
Furihata \cite{GEM2}, with the Los Alamos
version of the Quark-Gluon String Model code LAQGSM \cite{LAQGSM}
merged \cite{Mashnik02a}-\cite{fitaf} with  GEM2 \cite{GEM2},
as well as  with versions of the CEM2k and LAQGSM codes merged both
\cite{Mashnik02a}  with the
sequential-binary-decay code GEMINI by Charity \cite{GEMINI}.
The limited size of the present work does not allow us to discuss
these models here; description of the models may be found in the
original publications \cite{LAHET3}-\cite{GEMINI} and references therein.

Let us start with discussing results for the $^{129}$I target.
As we have done previously
(see, {\it e.g.,}~\cite{6,Titarenko04}),
we choose here one qualitative and one quantitative criterion to
judge how well our data are described by different models; namely,
the ratio of calculated cross section for the production of a given
isotope to its measured values
$\sigma^{cal} / \sigma^{exp}$ as a function of the mass number
of products (Fig.~1), and the mean simulated-to-experimental
data ratio (Table 1)
\begin{equation}\label{eqf}
\left<F\right>=10^{\sqrt{\left<(\log [ \sigma^{cal} / \sigma^{exp} ] )^2
\right>}}~,
\end{equation}
with its standard deviation~:
\begin{equation}\label{eqsf}
S\left(\left<F\right>\right)=10^{\sqrt{\left<\left(
\left|\log\left(
%\frac{\sigma_{cal}^i}{\sigma_{exp}^i}\right)
\sigma^{cal} / \sigma^{exp} \right)
\right|-\log(\left<F\right>)\right)^2\right>}}~.
\end{equation}

For such a comparison, out of all the 74 measured cross sections \cite{8},
only 42 were selected to satisfy some rules based on appreciation
of the physical principles realized in the models.
For instance, if only a long-lived isomer or only the ground state 
of a nuclide was measured, such nuclides were excluded from
the quantitative comparison, but if both were measured separately,
their sum was compared
with calculations. Such rules are essentially similar to those used
by Titarenko {\it et al.}~\cite{6,Titarenko04}.

To understand how different models describe nuclides produced
in the spallation  and fission or fragmentation regions,
we divided all 42 measured nuclides included in our
quantitative comparison into two
groups, spallation $(A \ge 95)$ and fission/fragmentation $(A < 95)$.
The left panel of
Tab. 1 shows values of  $\left<F\right>$ and $S\left(\left<F\right>\right)$
for all compared products (both spallation and fission/fragmentation),
while the right panel of this table
shows such results only for spallation;
$N$ is the total number of comparisons, $N_{30\%}$ is the number of
comparisons in which
calculated and measured values differ by not more than 30 \%, while
$N_{2.0}$
shows the number of comparisons
where the difference was not more than a factor of two.

We note that the codes CEM95 \cite{CEM95} and CEM2k \cite{CEM2k}
consider only competition
between evaporation and fission of excited compound nuclei
and calculate the fission cross sections
for a nuclear reaction on a heavy nucleus, but do not calculate the fission
fragments, as they do not contain a fission model.
The Bertini \cite{Bertini} and ISABEL \cite{ISABEL} INC's are used in our
calculations with the
default options of LAHET3 for evaporation/fission models; they
consider evaporation with the Dresner code \cite{Dresner} and a possible
fission of heavy compound nuclei using the Atchison RAL fission
model \cite{RAL}, but only if they are heavy enough $(Z > 71)$,
{\it i.e.,} they do not consider fission for such light targets
as $^{129}$I.

\begin{figure}
\vspace{-5mm}
\centerline{\hspace{-6mm} \epsfxsize 17 cm \epsffile{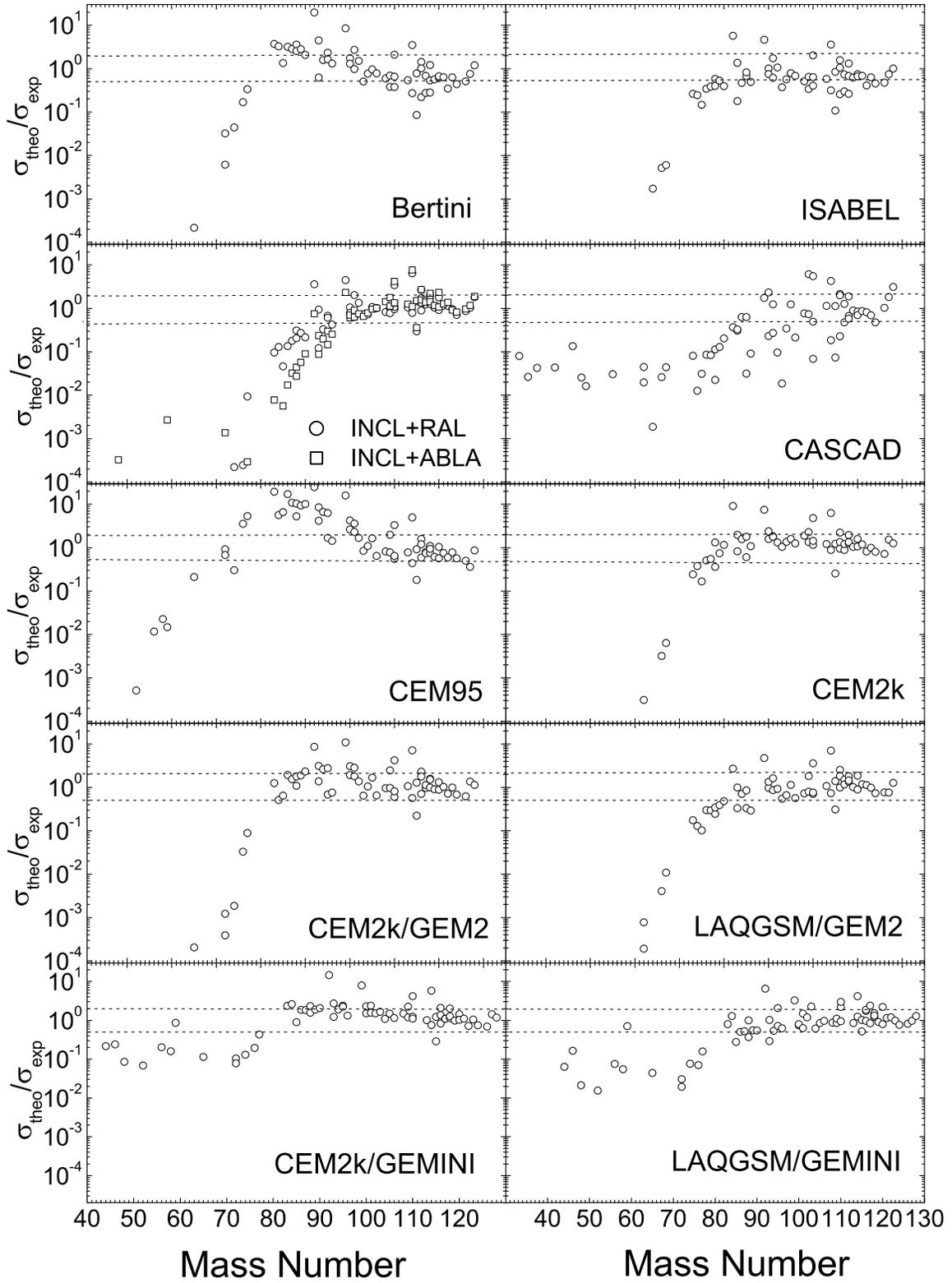}}
\vspace{-10mm}
\caption{
Ratio of theoretical to measured \cite{8}
cross sections of isotopes produced by a 660 MeV proton beam on a
15\% $^{127}$I + 85\% $^{129}$I target as a function of the product
mass number.
}
%\vspace{0.5in}
\end{figure}

{\noindent
CEM2k+GEM2 and LAQGSM+GEM2 consider fission using the
GEM2 model \cite{GEM2}
of only heavy nuclei, with $Z > 65$, {\it i.e.,} also not
considering fission of this target. Similarly, INCL+ABLA
\cite{INCL,ABLA} and CASCADE \cite{CASCADE} also do not consider
fission of $^{129}$I.
}
Only the code GEMINI by Charity \cite{GEMINI}
merged with CEM2k and LAQGSM considers fission (via
sequential binary decays) of practically all nuclei, and provides
fission products from this reaction.
This is why CEM2k+GEMINI and LAQGSM+GEMINI agree better than
all the other models tested here with experimental
data for this reaction, especially in the A = 40-80 mass region.

\begin{table}[h]
%\begin{table}[hbtp]
\centering
%\setcaptionwidth{4.00in}
\caption{Comparison of experimental and calculated results for all 42
selected product isotopes from $^{129}$I (left panel)
and for only 22 spallation products with A $\ge$ 95 (right panel)
}
\label{I129_all}
\vspace{0.5cm}
\begin{tabular}{|l|ccc||ccc|}
\hline
Model & \multicolumn{3}{|c||}{All 42 selected  isotope} &
 \multicolumn{3}{|c|}{22 spallation products with A $\ge$ 95}\\
\hline
 & $N/N_{30\%}/N_{2.0}$ & $\left<F\right>$ & $S\left(\left<F\right>\right)$
& $N/N_{30\%}/N_{2.0}$ & $\left<F\right>$ & $S\left(\left<F\right>\right)$ \\
    \hline
    Bertini+Dresner & 36/ 6/22 &  3.72 & 3.00  & 22/ 6/19 & 1.67 & 1.34\\
    ISABEL+Dresner  & 34/ 5/18 &  5.18 & 4.45  & 22/ 5/16 & 1.72 & 1.37\\
    INCL+Dresner    & 33/14/21 &  3.86 & 3.16  & 22/14/21 & 1.42 & 1.28\\
    INCL+ABLA       & 32/ 9/21 &  9.32 & 7.01  & 22/ 9/21 & 1.57 & 1.34\\
    CASCADE         & 42/ 9/15 & 11.05 & 5.19  & 22/ 9/14 & 3.32 & 2.75\\
    CEM95           & 40/10/20 &  5.40 & 3.52  & 22/ 9/18 & 1.78 & 1.44\\
    CEM2k           & 33/13/26 &  2.89 & 2.74  & 22/11/20 & 1.48 & 1.27\\
    LAQGSM+GEM2     & 33/13/22 &  3.16 & 2.68  & 22/13/21 & 1.50 & 1.34\\
    CEM2k+GEM2      & 35/10/28 &  5.03 & 5.04  & 22/ 8/20 & 1.60 & 1.35\\
    LAQGSM+GEMINI   & 42/12/29 &  4.28 & 3.58  & 22/17/21 & 1.31 & 1.21\\
    CEM2k+GEMINI    & 42/12/27 &  2.74 & 2.15  & 22/ 9/20 & 1.46 & 1.25\\
\hline
\end{tabular}
\end{table}

Newer calculations \cite{8} have shown that it is possible
to extend the fission model of GEM2 so that it describes also fission
of light nuclei, like $^{129}$I, and gives with CEM2k+GEM2 and LAQGSM+GEM2
for this reaction (as well as for other reactions, on other targets)
results very similar (even a little better)
to the ones provided by GEMINI
(compare the thin solid and dashed lines with the corresponding
thick lines in Fig. 2; Note that the lines here show the calculated total
yield of all products for every given mass number A, while the experimental
points show the yield of only measured (not all) isotopes, therefore,
the line should be, in general, higher than the experimental points).
For this, it is necessary to
fit the ratio of the level-density parameters for the fission
and evaporation channels, $a_f/a_n$. We think that it is possible to extend
in a similar way also the Atchison fission model \cite{RAL}
and the ABLA evaporation/fission model \cite{ABLA} to describe
fission of Iodine also with the Bertini+Dresner/Atchison,
ISABEL+Dresner/Atchison,
INCL+Dresner/Atchison, and INCL+ABLA options of LAHET3;
the same is true for the Dubna code CASCADE. Nevertheless,
we are not too optimistic about the predictive power of such
extended versions of these codes as they do not contain yet
reliable models for fission barriers of light nuclei.

To make the situation even more intricate, we note that when we
merge \cite{OurNewINC,Madeira04} CEM2k+GEM2 and LAQGSM+GEM2 with
the Statistical Multifragmentation Model (SMM) by Botvina
{\it et al.} \cite{SMM}, it is possible to
describe this reaction and get results very similar to the ones
predicted by CEM2k+GEMINI and LAQGSM+GEMINI without extending
the fission model of GEM2, {\it i.e.,} considering only INC, preequilibrium,
evaporation, and multifragmentation processes, but not fission of
$^{129}$I
(see the thick solid and dashed lines in Fig. 3).
We will discuss these and other similar
results in more details in a future publication.
Here, we note that it is impossible to make a correct
choice between fission and fragmentation reaction mechanisms
involved in our $p + ^{129}$I reaction by
comparing theoretical results with our (or other similar) measurements
of only product cross sections;
addressing this question would require analysis of two- or
multi-particle correlation measurements.

\vspace*{-10mm}

\begin{figure*}[h]
\begin{minipage}[h]{75mm}
\hspace{-9mm}
\includegraphics[width=75mm,angle=-90]{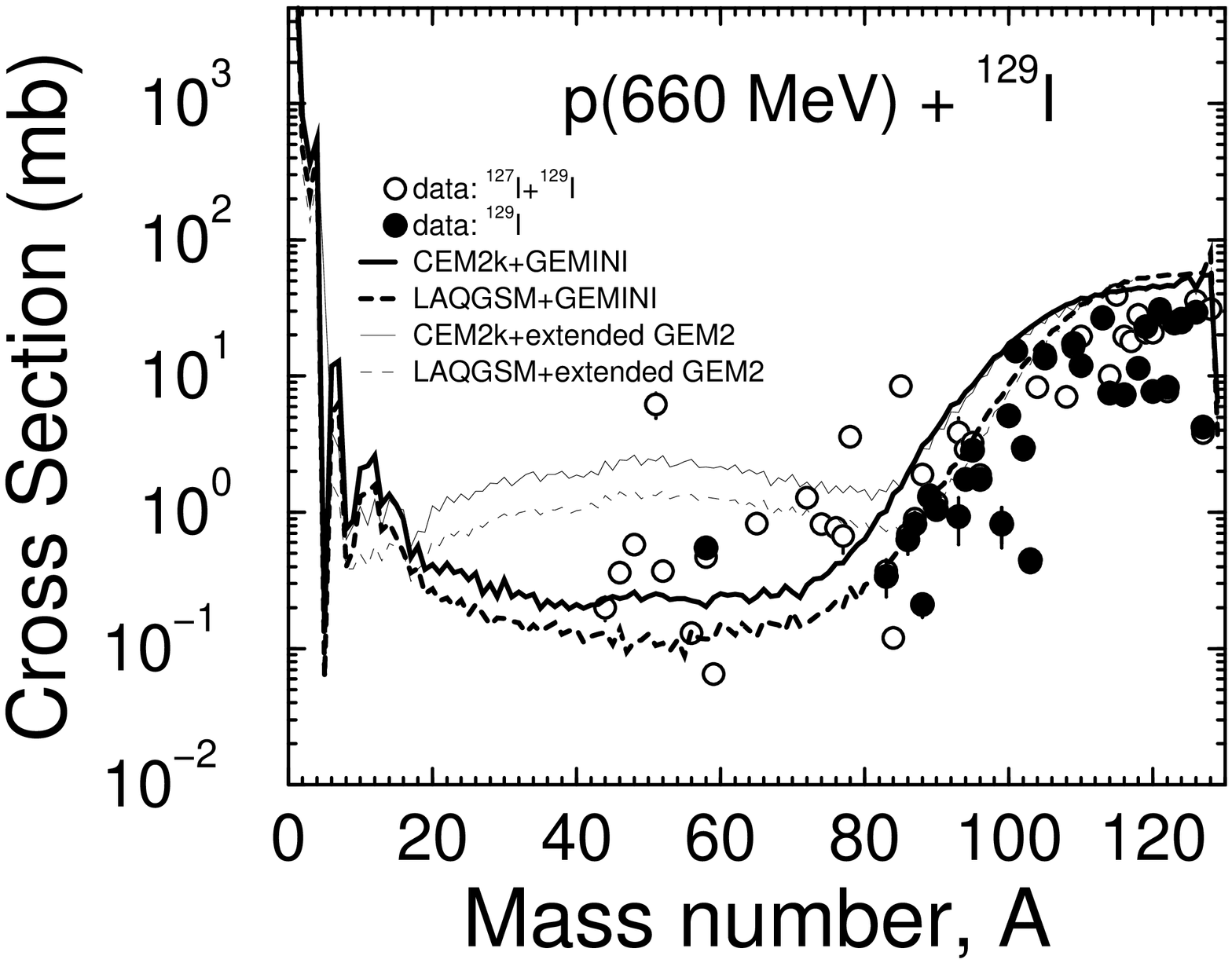}
\vspace{-5mm}
\caption{
Comparison of experimental and theoretical
cross sections for products from 660 MeV  p + I.
Open circles show data for
our 15\% $^{127}$I + 85\% $^{129}$I target,
while filled circles show estimated experimental data
for only $^{129}$I \cite{8}.
Thick solid and dashed lines show results for $^{129}$I
by CEM2k and LAQGSM
merged with GEMINI \cite{GEMINI}, while thin lines, corresponding
results by an extended version of GEM2 to consider fission
of light nuclei, as described in the text.
}
\end{minipage}
\hspace{\fill}
\begin{minipage}[h]{75mm}
\hspace{-9mm}
\includegraphics[width=75mm,angle=-90]{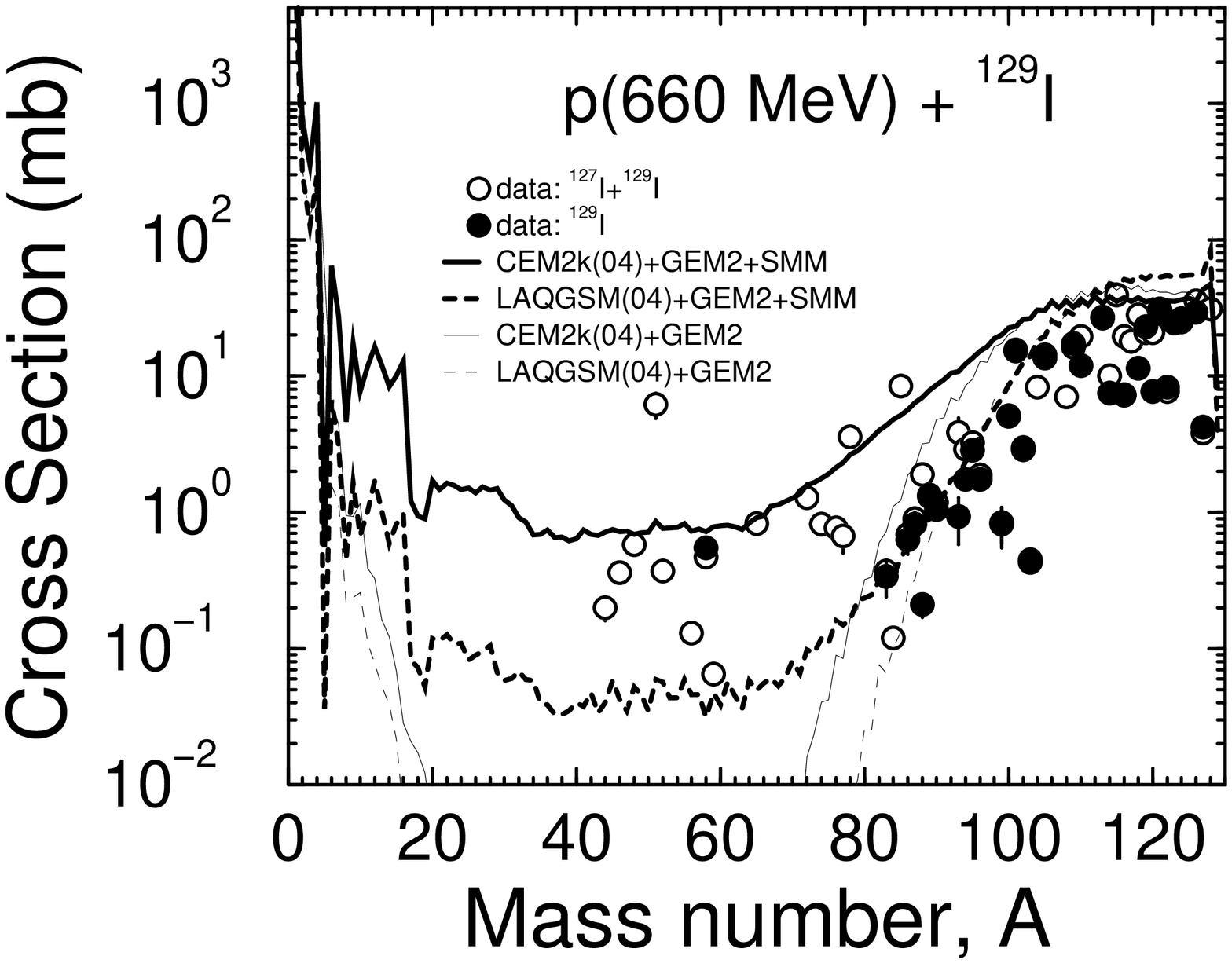}
\vspace{-5mm}
\caption{
The same experimental data as in Fig. 2 but compared with
results for $^{129}$I by CEM2k+GEM2 and LAQGSM+GEM2
(thin solid and dashed lines) and by these models
extended \cite{OurNewINC,Madeira04}
to consider also multifragmentation \cite{SMM}
of compound nuclei when their
excitation energy $E^*$ is above 2 MeV/A,
as a competitive channel
to evaporation of particles
after the preequilibrium stage of reactions
(thick solid and dashed lines); no fission is
considered here.
}
\end{minipage}
\end{figure*}

Results for the $^{237}$Np and $^{241}$Am targets are shown
in Figs. 4 and 5 and Tabs. 2 and 3, respectively. The situation
for these actinides is quite different from what we have
above for $^{129}$I. First, almost all
isotopes measured from these targets \cite{7} are fission products.
From all 53 measured isotopes from $^{237}$Np, only 37 are selected
for the quantitative comparison shown in Table 2,
with 32 of them being fission products; and from all 80 measured
isotopes from $^{241}$Am, only 45 are selected for the
comparison in Table 3,
and 44 of them are fission products. All codes tested here describe
fission of actinides
(we do not compare results by CEM95 and CEM2k with the Am and Np data,
as these codes do not calculate fission fragment production),
but the agreement of the calculations with these measured
data \cite{7} is much worse than we have for $^{129}$I, for all codes.
From Tabs. 2 and 3, we see that none of the codes have
a mean deviation factor $<F>$ less than a factor of two,
whether we compare both fission and spallation products (left panels
in Tabs. 2 and 3), or only the fission products (right panels).
A little better agreement with these data
in comparison with other codes are provided
by the old models Bertini+Dresner/RAL and ISABEL+Dresner/RAL from
the LAHET transport code. GEMINI, that works so well for $^{129}$I,
does not describe the Am and Np fission products well. This is not a
surprise, as GEMINI was developed to describe well sub-actinides;
it needs to be further

\begin{figure}
\vspace{-1.cm}
\centerline{\hspace{-8mm} \epsfxsize 16cm \epsffile{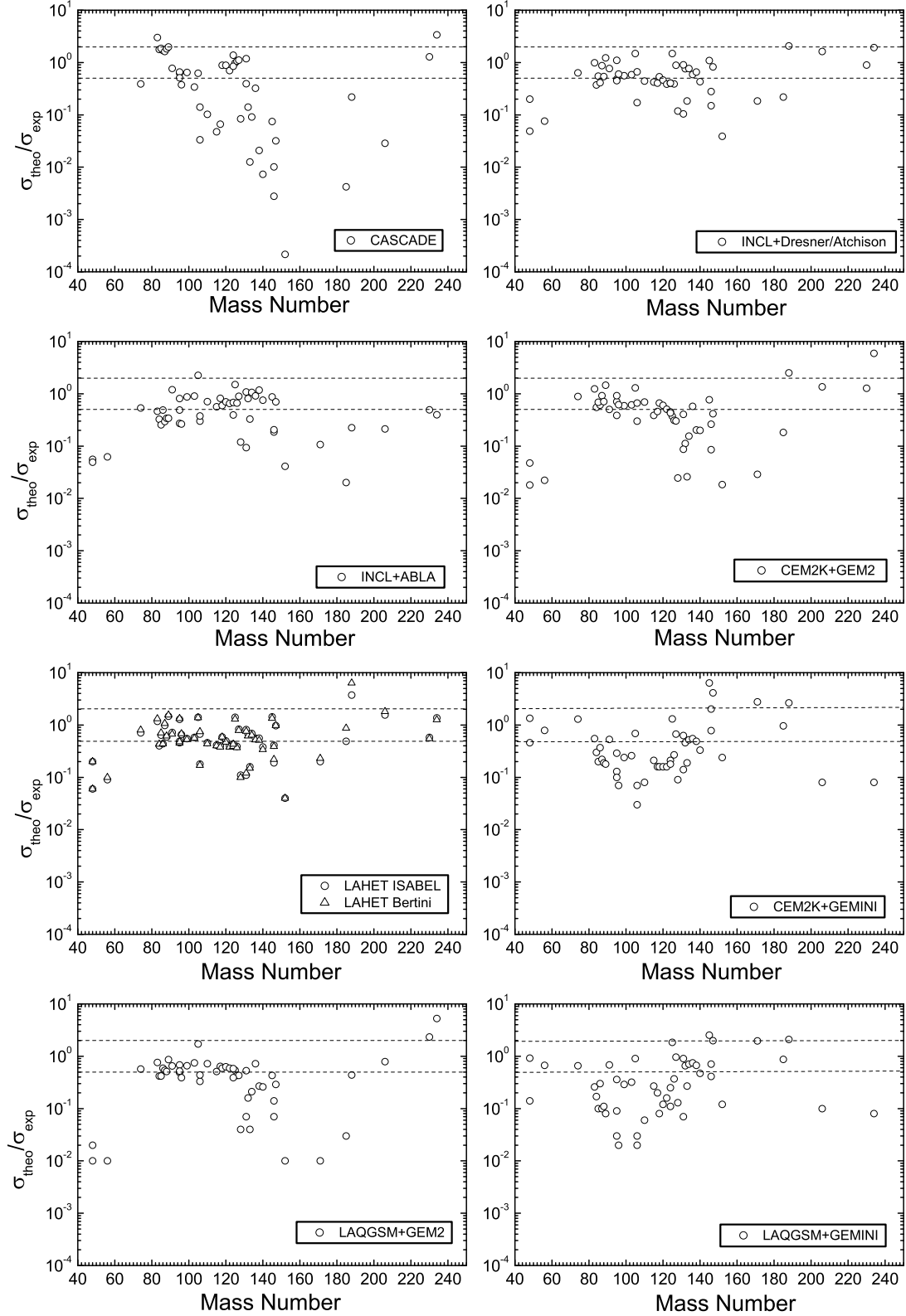}}
\vspace{-10mm}
\caption{
Ratio of theoretical to measured \cite{7}
cross sections of isotopes produced by a 660 MeV proton beam on a  $^{237}$Np
target as a function of the product
mass number.
}
\end{figure}

\begin{figure}
\vspace{-1.cm}
\centerline{\hspace{-8mm} \epsfxsize 16cm \epsffile{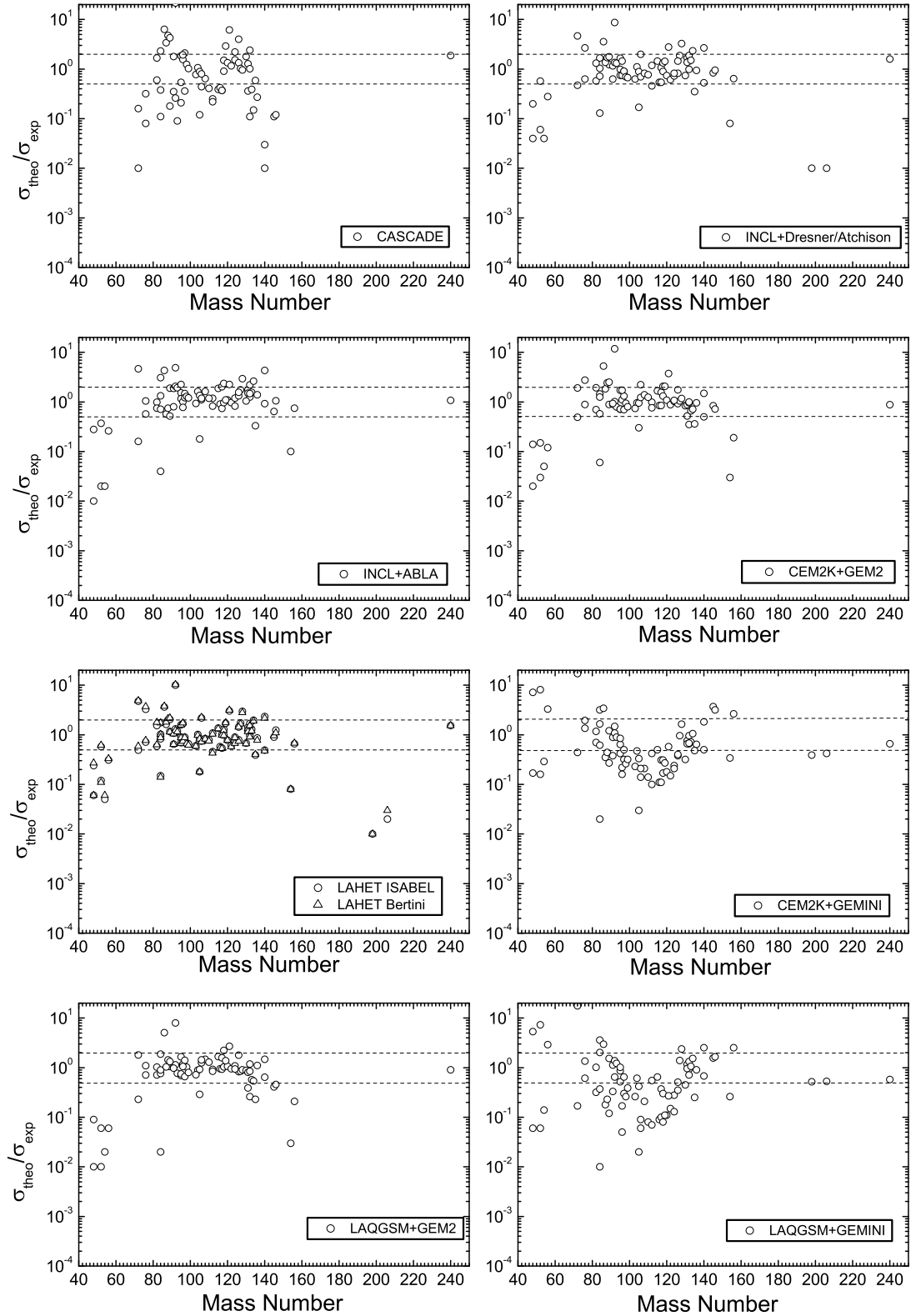}}
\vspace{-10mm}
\caption{
Ratio of theoretical to measured \cite{7}
cross sections of isotopes produced by a 660 MeV proton beam on a  $^{241}$Am
target as a function of the product
mass number.
}
\end{figure}

\noindent{
 improved to become a reliable tool also for actinides.
The newly developed code CEM2k and LAQGSM merged with GEM2 and INCL+ABLA,
that work so well for many other reactions, also fail to reproduce
these Np and Am data, indicating that these codes still have
deficiencies and need further improvement.

}

The qualitative comparisons shown in Figs. 4 and 5
are even more drastic: We see deviations between some calculated and
measured cross sections as large as two orders of magnitude, and
even higher.

\begin{table}[h]
\centering
%\setcaptionwidth{4.00in}
\caption{Comparison for all 37
selected isotopes (spallation and fission) from $^{237}$Np
(left panel) and  for only 32 selected isotopes (fission products) with
A $<$ 175 (right panel)}
\label{Np237_all}
\vspace{0.5cm}
\begin{tabular}{|l|ccc||ccc|}
\hline
 Model & \multicolumn{3}{|c||}{All 37 selected  isotope} &
 \multicolumn{3}{|c|}{32 fission products with A $<$ 175 }\\
\hline
 & $N/N_{30\%}/N_{2.0}$ & $\left<F\right>$ & $S\left(\left<F\right>\right)$
& $N/N_{30\%}/N_{2.0}$ & $\left<F\right>$ & $S\left(\left<F\right>\right)$ \\
    \hline
CASCADE             & 32/ 5/16 & 11.29 & 7.75  & 27/ 4/15 &  9.75 & 7.40 \\
  INCL+Dresner/RAL  & 37/ 8/19 &  3.35 & 2.39  & 32/ 7/16 &  3.51 & 2.49 \\
  INCL+ABLA         & 37/ 9/14 &  4.52 & 2.92  & 32/ 9/14 &  4.09 & 2.78 \\
 Bertini+Dresner/RAL& 37/ 6/20 &  3.18 & 2.26  & 32/ 4/16 &  3.29 & 2.30 \\
  ISABEL+Dresner/RAL& 37/ 5/19 &  3.18 & 2.26  & 32/ 5/16 &  3.34 & 2.37 \\
      CEM2k+GEM2    & 37/ 6/18 &  5.43 & 3.47  & 32/ 5/16 &  5.80 & 3.70 \\
      CEM2k+GEMINI  & 36/ 4/11 &  3.87 & 2.05  & 32/ 3/10 &  3.57 & 1.88 \\
      LAQGSM+GEM2   & 37/ 2/15 &  7.10 & 4.29  & 32/ 1/14 &  7.28 & 4.47 \\
      LAQGSM+GEMINI & 36/ 5/13 &  4.82 & 2.69  & 32/ 4/12 &  4.69 & 2.64 \\
\hline
\end{tabular}
\end{table}

%\newpage

\begin{table}[h]

\centering
%\setcaptionwidth{4.00in}
\caption{Comparison for all 45 selected isotopes (spallation
and fission) from $^{241}$Am (left panel) and for only 44 selected
isotopes (fission products) with A $<$ 175 (right panel)}
\label{Am241_all}
\vspace{0.5cm}
\begin{tabular}{|l|ccc||ccc|
}\hline
 Model & \multicolumn{3}{|c||}{All 45 selected  isotope} &
 \multicolumn{3}{|c|}{44 fission products with A $<$ 175 }\\
\hline
 & $N/N_{30\%}/N_{2.0}$ & $\left<F\right>$ & $S\left(\left<F\right>\right)$
& $N/N_{30\%}/N_{2.0}$ & $\left<F\right>$ & $S\left(\left<F\right>\right)$ \\
    \hline
CASCADE             & 41/ 7/16 &  6.43 & 4.10  & 40/ 7/15 &  6.57 & 4.15 \\
  INCL+Dresner/RAL  & 45/14/37 &  2.43 & 2.13  & 44/14/36 &  2.44 & 2.15 \\
  INCL+ABLA         & 45/17/36 &  2.93 & 2.68  & 44/16/35 &  2.96 & 2.69 \\
 Bertini+Dresner/RAL& 45/19/35 &  2.26 & 1.98  & 44/19/34 &  2.28 & 1.99 \\
  ISABEL+Dresner/RAL& 45/19/35 &  2.28 & 1.99  & 44/19/34 &  2.30 & 2.01 \\
      CEM2k+GEM2    & 45/22/33 &  2.73 & 2.48  & 44/21/32 &  2.76 & 2.50 \\
      CEM2k+GEMINI  & 45/ 7/18 &  3.14 & 2.01  & 44/ 7/17 &  3.18 & 2.02 \\
      LAQGSM+GEM2   & 45/22/34 &  3.34 & 3.12  & 44/21/33 &  3.38 & 3.14 \\
      LAQGSM+GEMINI & 45/ 8/23 &  3.64 & 2.51  & 44/ 8/22 &  3.69 & 2.53 \\
\hline
\end{tabular}
\end{table}

Let us mention that such a big disagreement between calculations
by the codes tested here and these experimental data does not
mean that the codes are in general too bad and they can not predict
well any data. The measurements \cite{7,8} were done by the gamma-spectrometry
method which allows us to measure only a small part of all products from
any reaction. In addition, most of the measured cross sections
are cumulative, while all codes provide only independent cross
sections that need to be summed up with the corresponding
branching ratios to be able to compare with the measured
cumulative yields. That makes such comparisons more difficult;
many products are measured either only in the ground
states or only as long-lived
isomers, and such data can not be compared with our
calculations (see details about this in \cite{6,Titarenko04}).
It is much easier and convenient to test codes against recent
measurements done at GSI in Darmstadt, Germany
using inverse kinematics
for interactions of $^{56}$Fe, $^{197}$Au, $^{208}$Pb, and $^{238}$U
at 1 GeV/nucleon and several lower energies with
liquid $^1$H, as well as with many heavier targets up to $^{208}$Pb.
References to most of the GSI measurements and
many tabulated experimental cross sections may be found on the Web
page of Prof. Schmidt \cite{SchmidtWebPage}.
The GSI measurements  provide
a very rich set of cross sections for production of practically
all possible isotopes from such reactions in a ``pure" form,
{\it i.e.}, individual cross sections from a specific given bombarding isotope
(or target isotope, when considering reactions in the usual kinematics,
p + A). Such cross sections are much easier to compare to models than the
``camouflaged" data from $\gamma$-spectrometry measurements,
like the ones analyzed here,
and many of the codes tested here describe very well most of the GSI
measurements, with a mean deviation factor usually not higher than
a factor of two
(see, {\it e.g.}, \cite{6,INCL,ND2001,GSI03} and references therein).
With these circumstances in mind, our present comparison of
the I, Np, and Am measurements \cite{7,8}
with calculations by different models does not
pretend either to be comprehensive
or to pick up ``the best" tested code. Rather,
our goal is to investigate problems some models may still have in
reproducing well some specific measurements, with a hope that this
would help the authors of codes to improve their models.
The results of the present work show
that all codes tested here still have some big problems in a correct
description of many of the $^{129}$I, $^{237}$Np, and  $^{241}$Am
data and all models should be further improved.

\noindent{\bf Conclusions}

We have analyzed the recent JINR (Dubna, Russia) measurements on nuclide
production cross sections from interaction of 660 MeV proton beams with
radioactive targets of enriched $^{129}$I (85\% $^{129}$I and 15\% $^{127}$I),
$^{237}$Np, and $^{241}$Am \cite{7,8} with eleven different models,
realized in eight transport codes and event-generators: LAHET (Bertini, ISABEL,
INCL+ABLA, and INCL+RAL options), CASCADE, CEM95, CEM2k, LAQGSM+GEM2,
CEM2k+GEM2, LAQGSM+GEMINI, and CEM2k+GEMINI. We found out that all
these models have problems in a correct description of many of these
cross sections, though some of these models describe very well most of the
recent measurements done at GSI  using inverse
kinematics, as well as many other reactions.
None of the tested here models is
able to reproduce well all the JINR data and all of them should be further
improved. Development of a better
universal evaporation/fission model should be of a highest priority.

In the case of the $^{129}$I target, products with mass numbers $A = 40-70$
were measured \cite{8}. Such isotopes can not be described by the
models tested here as spallation products, and may be considered
as produced either via fission or multifragmentation. The CEM2k and
LAQGSM codes merged with the sequential binary decay model GEMINI
can reproduce such isotopes as fission products. On the other hand,
CEM2k and LAQGSM merged with the Statistical Multifragmentation Model
(SMM) can also reproduce these isotopes via multifragmentation,
{\it i.e.}, considering only intranuclear cascade, preequilibrium,
multifragmentation, and evaporation processes,
without fission of $^{129}$I.
Similar results were obtained with these codes for p + $^{56}$Fe reactions
at 1.5, 1.0, 0.75, 0.5, and 0.3 GeV measured at GSI in inverse kinematics.
We conclude that it is impossible to make a correct choice
between fission and fragmentation reaction mechanisms analyzing
only measurements on product cross sections, at least for interactions
of intermediate-energy nucleons with medium-mass targets;
addressing this question would require analysis of two- or
multi-particle correlation measurements.

\indent
This work was partially supported by the US Department of Energy,
Moldovan-US Bilateral Grants Program, CRDF Project MP2-3045-CH-02,
and NASA ATP01 Grant NRA-01-01-ATP-066.

\end{document}